\def\junk#1{}

\documentclass[usletter, 10pt, conference]{ieeeconf}
\usepackage{setspace} 
\usepackage[usenames,dvipsnames,svgnames,table]{xcolor}
\usepackage{graphics} 
\usepackage{epsfig} 
\usepackage[caption=false,font=footnotesize]{subfig}
\usepackage{times} 
\usepackage[cmex10]{amsmath}
\usepackage{amssymb}
\usepackage{amstext}
\usepackage{url}
\usepackage{algorithm}
\usepackage{algorithmic}
\usepackage{mdwmath}
\usepackage{mdwtab}
\usepackage{epstopdf}
\usepackage{multirow}

\usepackage[switch]{lineno}

\usepackage{verbatim}

\usepackage[normalem]{ulem} 

\newtheorem{definition}{Definition}

\newtheorem{remark}{\textbf{Remark}}
\newtheorem{example}{\textbf{Example}}

\author{
\begin{tabular}{cccc}
\vspace{-0.05in}
Saber Mirzaei &&   Flavio Esposito   \\
{\small \; smirzaei@cs.bu.edu } && {\small \; fesposito@exegy.com } \\
\small{Computer Science Department} && {\small Exegy Inc. }  \\
{\small  Boston University, MA} && \small{ St. Louis, MO}
\end{tabular}
}

\title{\Large \bf
An Alloy Verification Model for Consensus-Based Auction Protocols}

\begin{document}


\maketitle
\thispagestyle{empty}
\pagestyle{empty}


\begin{abstract}
%
Max Consensus-based Auction (MCA) protocols are an elegant approach to establish conflict-free distributed allocations in a wide range of network utility maximization problems. 
A set of agents independently bid on a set of items, and exchange their bids with their first hop-neighbors for a distributed (max-consensus) winner determination.
MCA protocols have been proposed, $e.g.$, to solve the task allocation problem for a fleet of unmanned aerial vehicles, in smart grids, or in distributed virtual network management applications.
%
Misconfigured or malicious agents participating in a MCA or an incorrect combination of policy instantiations can lead to oscillations of the protocol,  causing, $e.g.$, Service Level Agreement (SLA) violations.
%

In this paper we propose a formal, machine-readable, Max-Consensus Auction model encoded in the Alloy lightweight modeling language. The model consists of a network of agents applying the MCA mechanisms instantiated with potentially different policies, and a set of predicates to analyze its convergence properties.
%
We were able to verify that even when all agents follow the protocol, MCA is not resilient against rebidding attacks, and that the protocol fails (to achieve a conflict-free resource allocation) for some specific combinations of policies.
%
Our model can be used to verify, with a ``push-button" analysis, the convergence of the MCA mechanism to a conflict-free allocation under a wide range of policy instantiations. 




    \end{abstract}
\vspace{-0.1in}

\section{Introduction}
    \label{sec:intro}
    %
Resource allocation problems are ubiquitous in distributed systems. The Max Consensus-based Auction (MCA) protocol is a recent approach that allows a set of communicating agents to rapidly obtain a conflict-free (distributed) allocation of a set of items, given a common network utility maximization goal. Without calling it MCA, recent work~\cite{cbba,CADpaper} demonstrated how max-cosnensus auction protocols provide desirable performance guarantees with respect to the optimal network utility. The MCA protocol consists of two mechanisms: a bidding mechanism, where agents independently bid on a single or on multiple items,  and an agreement (or consensus) mechanism, where agents exchange their bids for a distributed winner determination.

The use of MCA protocols was proposed to solve resource allocation problems across several disciplines.
To our knowledge, its first use appeared to solve the distributed task assignment problem~\cite{cbba}, where a fleet of unmanned aerial vehicles bid to assign a set of tasks (geo-locations to be covered.)
MCA protocols were also proposed  for distributed virtual network management applications~\cite{CADpaper}, where federated infrastructure providers bid to host virtual nodes and virtual links on their physical network, in attempt to embed a wide-area cloud service.
More recently, MCA protocols have been  also proposed to solve the economic dispatch problem in a distributed fashion, $i.e.$, the problem of allocating  power generation tasks among available units in a smart-grid~\cite{MCA-grid}.

Each (invariant) mechanism of the MCA protocol may be instantiated with different policies. An MCA policy is a variant aspect of the bidding or the agreement mechanism, and represents an high-level application goal. Examples of policies for the bidding mechanism are the (private) utility function used to generate bids, or the number of items on which agents simultaneously bid on, in each auction round. Note that MCA does not require a centralized auctioneer~\cite{cbba,CADpaper}.~\footnote{Variation of policies may induce different behavior. For example, second price auctions on a {\it single} item are known to have the strong property of being truthful in dominant strategies, $i.e.$, the auction maximizes the revenue of the bidders who do not have incentives to lie about their true (utility) valuation of the item. In the MCA settings however, truthful strategies may not work as there is uncertainty on whether more items are to be assigned in the future; bidders may have incentives to preserve resources for stronger future bids.}
%

Earlier work on protocols verification  established how certain combinations of policy instantiations may lead to incorrect  behaviors of a protocol~\cite{arye2011toward, zave2009lightweight}. Similarly, in this paper we analyze the convergence properties of the MCA protocol  under various settings using a lightweight, machine-readable, Alloy~\cite{Jackson:2006} verification model.
Our aim is to show how certain combinations of MCA policies, obtained by design, resulting from misconfigured or malicious agents, may break the convergence of the MCA protocol causing the application to fail and inducing $e.g.$, Service Level Agreement (SLA) violations, energy inefficiencies, or the loss of expensive unmanned vehicles (whose software may fail under an MCA instability.)
By MCA convergence, we mean the attainment of a distributed conflict-free assignment of the items on auction.

%
In particular, we present the following contributions: ($i$) we identify the common mechanisms of several existing max-consensus auction protocols, renaming MCA such unifying set of mechanisms, and we separate them from their policies in our Alloy model. We then verify the impact of some of the policy combinations on correctness of the protocol.
($ii$) We describe the Max-Consensus Auction mechanism, and some applications to motivate its versatility in Section~\ref{sec:problem}. As a case study, we dissect one particular application: the distributed virtual network mapping problem (defined in Section~\ref{sec:problem}), $i.e.$, the NP-Hard problem of assigning, or mapping, constrained virtual nodes and virtual links (items) to physical nodes (agents) and loop-free physical paths,  belonging to multiple federated infrastructure providers.  ($iii$) In Section~\ref{sec:alloy} we overview the basic concepts of the Alloy Modeling Language and the Alloy Analyzer in context with our model, described in Section ~\ref{sec:VNE-model}, and available at~\cite{BUalloy}. The model consists of a network of agents together with the set of rules used to asynchronously resolve conflicts, and a set of predicates to analyze the convergence property, when agents are instantiated with different MCA policies. 
($iv$) In Section~\ref{sec:property-checking} we present the analysis of the convergence properties of the MCA protocol. In particular, we present a set of counter-examples to show how the MCA protocol fails to reach a conflict-free assignment of the items on auction, for a particular combination of policy instantiations (Section~\ref{sec:property-checking}.)
 We finally discuss some relevant work in Section~\ref{sec:related-work} and conclude our paper in Section~\ref{sec:conclusions}.

\junk{
\textcolor{red}{Our main contribution, the Alloy model of the consensus-based virtual network mapping, is presented in Section~\ref{sec:VNE-model}.
Virtual network mapping problem is one of the many applications of MCA that we have chosen as our case study to model using Alloy.
In this section the model is explained while divided into two logical parts, static and dynamic sub-models. The static part models the underlying networks (hosting physical network and the target virtual network to be mapped on it), while the dynamic sub-model captures the state transition of the system (based on the an MCA protocol).
Both the static and the dynamic models of the MCA protocol enable us to study the behavior and some properties of the MCA protocols (such as the conversion of the transition system under different settings of the protocol).}
 The model consists of a network of agents together with the set of rules used to asynchronously resolve conflicts, and a set of predicates to analyze the convergence property when agents are instantiated with different MCA policies. Using our model,
 we present a set of counter examples to show how MCA fails to reach a conflict-free assignment of the items on auction, for a particular combination of policy \sout{instantiations} \textcolor{red}{configuration (to many use of the word instantiate)} (Section~\ref{sec:property-checking}).
}
Once released, both our static and  dynamic models will serve as a baseline tool for a deeper investigation of the MCA convergence properties, when the bidding agents are instantiated with possibly conflicting policies.
%




\junk{
- Start with mentioning how important is netowork virtualization, introducing alloy layter:
- Some of the misconfigurations are due to poor design and ``agile" development requirement, with poor protocol design.
- Describe the importance of having a correct protocol cite the yankee group report that says that more than $60\%$ of outages comes from network management, and more than $50\%$ of the cost comes from IT.
- Define the context: We call service providers the players that do not own the infrastructure but provide a (cloud-based) service. Infrastructure providers own instead the physical network resources. A cloud provider can be a lessor or a lessee of the network infrastructure, and can act as both service and infrastructure provider.
- Alloy is a tool that can potentially limit such outages, therefore saving its costs to cloud provider.
- Define the contributions of this paper referrin to the outline of this paper
}

\section{The Max-Consensus Auction Protocol}
    \label{sec:problem}
In this section, we first introduce the Max Consensus-based Auction mechanisms, and then we describe few motivating applications on which such a protocol may be,  or was already applied, with particular attention to the virtual network mapping application, that we use as a case study for the rest of the paper. 

\vspace{-1mm}
\subsection{The Mechanisms}\label{sec:max-consensus}
\vspace{-1mm}

Consider a set $\mathcal{I}$ of independent agents (or nodes), that need to allocate in a distributed fashion a set $\mathcal{J}$ of items. Each agent is associated with a private utility $\mathbf{u}_i \in \mathbb{R}_{+}^{|\mathcal{J} |}$, that represents the benefit (or cost) of hosting an element of  $\mathcal{J}$.
As in~\cite{cbba} and \cite{CADpaper}, we assume that agents cooperate to reach a Pareto optimal solution: $\sum_{i \in \mathcal{I}} u_i$.  
A Max-Consensus Auction protocol consists of two independent mechanisms: $(i)$ a bidding mechanism, where agents independently bid on the items in $\mathcal{J}$, and $(ii)$ an agreement mechanism, where bids are exchanged with the logical neighbors for a distributed winner determination. In particular, an asynchronous agreement is sought on the {\it maximum} bid on each item to be assigned.

During the bidding phase, using their (private) utility function $\mathbf{u}$, each agent independently assigns bid values on a subset of $\mathcal{J}$.  
Each agent constructs a vector {\bf b}, where ${\bf b}_{ij}$ is the bid of agent $i$ on item $j$. The utility function ${\bf u}_i$, used to generate the bids, may depend also on previous bids. Agents have a limited budget, $i.e.$ the capacity of physical node to host virtual nodes, and their bids on current items depend on how many items they have won in the past. Formally, the max-consensus on a set of items is defined as follows~\cite{consensusBook}:
\begin{definition}\label{def:maxConsensus}({\it max-consensus}).
Given a network of agents $G$, composed by a set of agents $\mathcal{I}$, an initial bid vector of nodes  $\mathbf{b}(0) \stackrel{\Delta}{=} (\mathbf{b}_1(0), \ldots, \mathbf{b}_{|\mathcal{I}|}(0))^{\rm |\mathcal{J}|}$ \footnote{The notation implies that each vector $\mathbf{b}_i(0)$ has size $|\mathcal{J}|$.}, a set of neighbors $\mathcal{N}_i \; \forall  i \in \mathcal{I}$,  and the consensus algorithm for the communication instance $t+1$:
\vspace{-2 mm}
\begin{equation}
\vspace{-1mm}
\mathbf{b}_i(t+1) = \max_{j \in \mathcal{N}_i  \cup \{ i \} }\{\mathbf{b}_j(t)\} \;\; \forall i \in \mathcal{I},
\vspace{-1mm}
\end{equation}
{\it Max-consensus} on the bids among the agents is said to be achieved with convergence time $\tau$, if $\exists \; \tau \in \mathbb{N}$ such that $\forall \; t \geq \tau$ and $\forall \; i, i^{\prime} \in \mathcal{I}$,
\vspace{-2 mm}
\begin{equation}\label{maxconsensus}
\vspace{-1mm}
\mathbf{b}_i(t) = \mathbf{b}_{i^{\prime}}(t) = \max\{  \mathbf{b}_1(0), \ldots, \mathbf{b}_{|\mathcal{I}|}(0)  \},
\vspace{-1mm}
\end{equation}
\junk{
\textcolor{red}{
\begin{equation}\label{maxconsensus}
\vspace{-1mm}
{\mathbf{b}_i(t) = \mathbf{b}_{i^{\prime}}(t) = \max\{  \mathbf{b}_1(\tau), \ldots, \mathbf{b}_{|\mathcal{I}|}(\tau) \},
\vspace{-1mm}
\end{equation}
}
}
}

\noindent
where ${\rm max\{\cdot\}}$ is the component-wise  maximum.
\end{definition}


During the bidding phase, agents also save the identity of the items in a bundle vector
$\mathbf{m}_i \in \mathcal{J}^{T_{i}}$, where  $T_{i}$ is the target number of items that can be assigned to agent $i$, and a vector of time stamps ${\bf t}$, $i.e.$, the time at which the bid was generated. The bid generation time-stamps are used in the agreement phase to resolve assignment conflicts in an asynchronous fashion; when transmitted among agents, bids can in fact arrive out of order.
%
%
After the bidding phase, each physical node exchanges the bids with its neighbors, updating  an assignment vector $\mathbf{a}_i \in \mathcal{I}^{|\mathcal{J}|}$ with the latest information on the current assignment of all items.


%
   \begin{figure}[t]
\begin{center}
    \includegraphics[scale=0.32]{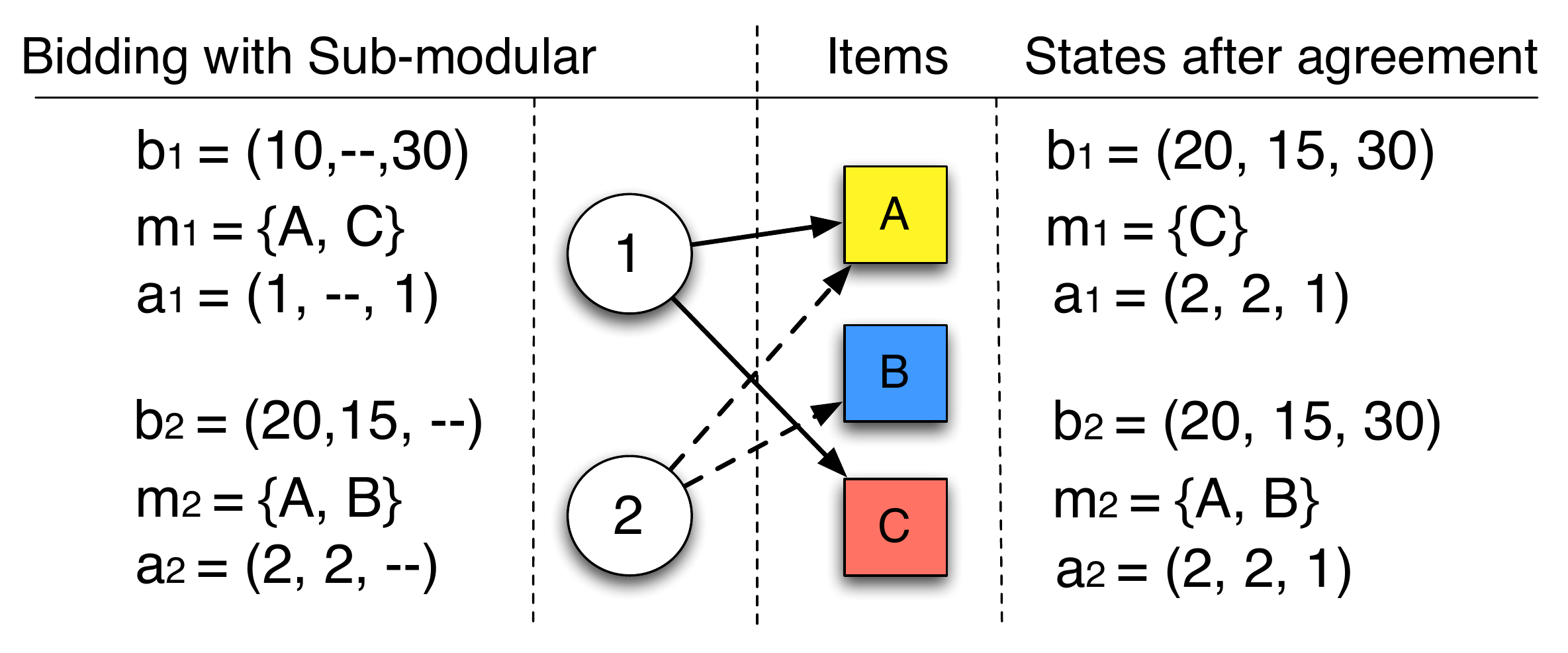}
\end{center}
\vspace{-0.2in}
\caption{Two agents (1,2) independently bid on three items (A,B,C), and exchange their bid and allocation vectors applying a distributed max-consensus auction protocol.}
\label{fig:example}
\vspace{-0.2in}
\end{figure}

 {\it
\begin{example}\label{example}
Consider Figure~\ref{fig:example}: agents $1$ and $2$ independently bid on three items $(A,B,C)$. Agent $1$ assigns a value of $10$ on item $A$, and a value of $30$ on item $C$, $b_1 = (10,-, 30 )$; then agent $1$ stores the identity of items A and C  in its bundle vector, $m_1 = (A,C)$, and assigns itself as a winner for both items, $i.e.$, sets the allocation vector $a_1$ with its own identifier for items $A$ and $C$. The bidding phase of agent $2$ is similar. After bidding, the agents exchange their bids and allocation vectors. Agent $1$ learns that there is a higher bid for item $A$, and stores such higher bid in its bid vector, and the identity of the overbidding agent $2$ in its allocation vector (see Figure~\ref{fig:example}, right column). The protocol has reached consensus.  An additional agent $3$, connected to agent $1$ but not agent $2$, would receive the {\it maximum} bid so far on each item, as well as the latest allocation vector ${\bf a} = {\bf a}_1 =  {\bf a}_2$.
\end{example}
}

In Example~(\ref{example}), an agreement is found after the first bidding phase. In general, for more elaborate (non-fully connected) networks of agents, the mechanism iterates  over multiple node bidding and agreement (consensus) phases.
 %
Note how a successful distributed allocation needs to be conflict-free, $i.e.$, the protocol can only assign each item to a single agent, while agents may win multiple items.

%

\begin{remark} {\it {\bf (no-rebidding allowed on lost items).}} \label{rm:overbid}
 A necessary condition to reach an agreement in an MCA protocol is that agents do not bid again on items on which they were overbid in previous auction rounds.
\end{remark}
\begin{remark} {\it {\bf (rebidding is convenient on items subsequent to an outbid.)}} \label{rm:rebidding}
Note how bids generated subsequently to an outbid item are outdated, because they were generated assuming a lower budget.
Hence, agents may rebid assigning a higher utility to the items already in their bundle, but subsequent to an outbid item.
\end{remark}
\begin{remark} {\it {\bf (sub-modularity of the bidding function).}} \label{rm:sm}
%
Assigning a set of items to a set of agents is equivalent to a Set Packing Problem, which is NP-hard~\cite{karp}.  Earlier work~\cite{cbba, CADpaper} has shown that if agents generate their bids using a sub-modular  function {\bf u}, then the network utility cannot be arbitrarily low. In particular, in~\cite{CADpaper} it has been shown that the allocation resulting from the MCA protocol has an approximation ratio of  $(1-\frac{1}{e})$ with respect to the optimal network utility $\sum_i {\bf u}_i$.

In the context of the MCA protocol, a sub-modular function is defined as follows:



%
\begin{definition}\label{def:submodularity} ({\it sub-modular function.})
The marginal utility function $u(j, {\bf m})$ obtained by adding an item $j$ to an existing bundle {\bf m}, is sub-modular if and only if
\vspace{-1 mm}
\begin{equation}
u(j, {\bf m^{\prime}}) \geq  u(j, {\bf m}) \; \forall \; {\bf m^{\prime}} \subset {\bf m}.
\end{equation}
\end{definition}
If an agent uses a sub-modular utility function, a value of a particular item $j$ cannot increase because of the presence of other items in the bundle ${\bf m_i}$ (the data structure keeping track of the items currently assigned to bidder $i$).
This implies that {\it bids on subsequent items cannot increase}. An example of sub-modular utility function is the residual capacity of a physical node bidding to host virtual nodes;  the residual (CPU) capacity can in fact only decrease as virtual nodes to be supported are added to the bundle vector {\bf m}.
\end{remark}

\vspace{-2mm}
\subsection{MCA Case Study: the Virtual Network Mapping Problem}\label{sec:applications}
\vspace{-1mm}

MCA protocols have been used across a wide range of networked applications, (see, $e.g.$,~\cite{MCA-grid,cbba,CADpaper}). In this subsection, we define  the virtual network mapping problem, the application that we have chosen as a case study for our MCA Alloy model.  


\junk{
\noindent
{\bf Application 2: Mission Allocation Problems.} The trend of employing drones to replace human activities has increased, but the use of autonomous robotic agents to accomplish complex missions, such as Unmanned Aerial Vehicles (UAVs) and autonomous ground rovers, has been around for several years; see, $e.g.$~\cite{Bellingham01multi-taskallocation}.
Teams of heterogeneous UAVs are regularly employed in complex missions including intelligence, surveillance and reconnaissance operations~\cite{air-force-tr}, and a proper coordination of tasks among a fleet of robotic agents is of critical interest for a successful mission operation. For those applications, consensus-based auction mechanisms have already been employed~\cite{cbba}: robotic agents bid on tasks, $e.g.$, locations to visit, and autonomously clear the task assignment minimizing a notion of cost.
}

\junk{
\noindent
{\bf Application 3: Cloud Resource Allocation Problems.}
Network virtualization has recently spurred interest in both the business and the research communities:  via wide-area virtual network testbeds as GENI~\cite{geni}, the networking communities can concurrently experiment with new Internet architectures and protocols, each running on an isolated instance of the physical network. From a market perspective, this paradigm is appealing as it enables multiple infrastructure and service providers to experiment with new business models that range from leasing their infrastructure to hosting multiple concurrent network services.
A fundamental protocol that providers need to run to host different virtual networks is the virtual network mapping.
A virtual network is a set of virtual instances spanning a set of physical resources, $e.g.$, processes and physical links, and by network service we mean the commodity supplied by the slice, $e.g.$ an online game or the access to a distributed virtual network testbed. 
}
Given a virtual network (VN)  $H = (V_H,E_H, C_H)$ and a physical network  $G = (V_G,E_G, C_G)$,
where $V_H$ and $V_G$ are the sets of virtual and physical nodes, respectively, and $E_H$ and $E_G$ are the set of  virtual and physical links, respectively. Each node or link $e$ (physical or virtual) is associated with a capacity constraint $C(e)$.~\footnote{Each $C(e)$ could be a vector $(C_1(e),\ldots, C_\gamma(e))$ containing  different types of constraints, $e.g.$ physical geo-location, delay, or jitter.} The  virtual network mapping is the problem of finding  {\it at least} a mapping of $H$ onto a subset of $G$, such that each virtual node is mapped onto exactly one physical node, and each virtual link is mapped onto at least one loop-free physical path $p$ while maximizing some utility or minimizing a cost function.
Formally, the mapping is a function
%
     $\mathcal{M}: H \rightarrow (V_G, {\cal P})$
%
where ${\cal P}$ denotes the set of all loop-free paths in $G$.     $ \mathcal{M}$ is called a {\it valid mapping} if all constraints of $H$ are satisfied, and for every  $l^H = (s^H, r^H) \in E_H$, exists at least one physical loop-free path $p: (s^G,\ldots, r^G) \in {\cal P}$ where $s^H$ is mapped to $s^G$ and  $r^H$ is mapped to $r^G$.
The MCA protocol is not necessarily applied to virtual links, as physical nodes (infrastructure provider processes) can merely bid to host virtual nodes, and later run $k$-shortest path to map the virtual links.

\begin{remark}
In the rest of the paper, our notation refers to the VN mapping problem, but our verification results are independent from the application running on top of the MCA protocol. Adapting to other MCA applications merely requires a change of variable names.
\end{remark}


\section{Alloy Overview}
    \label{sec:alloy}

\junk{ SAD vs MAD policy
fact {
	some p: pnode |
		#(p.initBidTriples.bidTriple_w - NULL) < 1
		and
		(all bv:bidVector | p = bv.bvPn implies #vnodesWonBy[bv] < 1)
}
}

In this section we describe how the Alloy~\cite{Jackson:2006} language and analyzer work, and we give few examples of primitives that we used to model the MCA protocol.

\noindent
{\bf What is Alloy and how does it work?} The term ``Alloy" refers to both a formal language and an automated analyzer.
 The {\it Alloy Analyzer} translates the user models into SATs, $i.e.$, boolean satisfiability problems.
The Alloy language is a declarative specification language for modeling complex structures and behaviors in a system. Alloy is based on  first order logic, and is designed for model {\it enumeration}.
%
The Alloy language is is based on the notions of
relations and sets. For example, a physical link can be modeled with a relation between two members of a physical node set. A relation is a particular set whose members are tuples with a specific arity.

To verify the satisfiability of the SAT representing the model, the Alloy Analyzer uses a constraint solver~\cite{torlak2007kodkod}.
%
%
Checking satisfiability of a large SAT instance may be intractable (or in general, time consuming);
In general, checking the satisfiability of the translated SAT instance is exponential (unless $P=NP$)~\cite{cook1971complexity}, however, the scope of the Alloy Analyzer can be customized and limited to ensure termination of the checking process in a timely fashion.

\noindent
{\bf How can we build a model using the Alloy language?}
The Alloy language is {\it lightweight}, and shares standard features and elements with most programming languages, $e.g.$, modules and functions; some features have been instead introduced by  Alloy, such as the concepts of {\it signature} or {\it scope}~\cite{Jackson:2006}.
A {\it signature} declaration is denoted by the keyword \texttt{sig}, and models the sets of elements of the system.
For example when the MCA protocol is applied to solve a virtual network mapping problem, a basic signature for a physical node with a given hosting CPU capacity and some capacitated connections  with other physical nodes can be modeled as follows:

\vspace{-1.5 mm}
{\scriptsize
\begin{verbatim}
    sig pnode{
        pcp: one Int, // Physical Cpu Capacity
        id: one Int, // ID of pnode
        pconnections: Int -> pnode // set of Connections
    }
\end{verbatim}
}
\vspace{-1.5 mm}

Signatures may contain some properties that model relations between elements. For example, the signature {\tt pnode}  has three relations, two binary ({\tt pcp} and {\tt id}), as they relate two signatures, and one ternary relation. 
The Alloy language also allows us to express constraints on sets and relations. Such constraints are defined with  {\it constraint paragraphs}, labeled by the keyword \texttt{fun}, as in function, $i.e.$, a reusable expression that always outputs a relation, \texttt{pred}, as in predicate, whose output is always a boolean, and \texttt{fact} $i.e.$, a constraint valid for any instance of the model.

For example, to impose the constraint of non negative physical links  capacity, in our model we define a fact \texttt{positiveCap} as follows:

%
\vspace{-1.5 mm}
{\scriptsize
\begin{verbatim}
    fact positiveCap{
        all n:pnode | (n.pconnections).pnode >= 0
    }
\end{verbatim}
}
\vspace{-1.5 mm}
\noindent
where the operator ``\texttt{.}" is the inner join in relational algebra.

To check that the model satisfies specific properties, the Alloy language supports {\it assertions}.
Assertions are labeled with the keyword \texttt{assert}. 
For example, to assert that two disjoint agents (physical nodes) \texttt{n1} and \texttt{n2} have non equivalent identifiers, we use the following assertion:

\vspace{-1.5 mm}
{\scriptsize
\begin{verbatim}
    assert uniqueID{
        all disj n1, n2: pnode | n1.id != n2.id
    }
\end{verbatim}
}
\vspace{-1.5 mm}

\noindent Note that assertions are not enforced rules,  but merely properties that we are interested in verifying.

The Alloy language also supports {\it commands}, $i.e.$, calls to the Alloy Analyzer. 
For example, to verify whether an assertion holds on a previously defined model,  within a user-defined scope, we use the command \texttt{check}. The command \texttt{run} instead instructs the Alloy Analyzer to find a satisfying instance of the SAT of the model.
To check if the assertion \texttt{uniqueID} holds in all instances of a model scope containing up to three physical nodes, we run the following Alloy command:
 {\tt    \normalsize{ check uniqueID for 3}}.

\vspace{-2mm}
\section{The Alloy Model for Consensus-based Virtual Network Mapping}
    \label{sec:VNE-model}
In this section we overview our  MCA Alloy model, applied to the virtual network mapping problem  (defined in Section~\ref{sec:applications}.)
Our Alloy code is logically divided into a static and a dynamic sub-model: the static sub-model refers to the underlying hosting physical network, and the virtual nodes to be mapped with the max-consensus based auction protocol, while the dynamic sub-model captures the state transitions.

\noindent
{\bf Static Model.} 
A simplified version of the signature \texttt{pnode} was explained in Section~\ref{sec:alloy}. We extend this signature to include some biding policies and other ternary relations:

\vspace{-2 mm}
{\scriptsize
\begin{verbatim}
    sig pnode{
        pcp: one Int,
        pid: one Int,
        initBids: vnode->Int,
        initBidTimes: vnode->Int,
        pconnections: some pnode,
        p_T: one Int,
        p_u: one utility,
        p_RO: one release_outbid
        // add your policy here
    }
\end{verbatim}
}
\vspace{-2 mm}

In this signature, \texttt{initBids} models the initial values that an agent (physical node) assigns when bidding on a subset of items (a virtual node is modeled as \texttt{vnode}), and the ternary relation \texttt{initBidTimes} models the corresponding bidding time on the virtual nodes, used for the  MCA asynchronous conflict resolution mechanism.
The bidding policies are modeled using the binary relations $e.g.$, ${\tt p\_T}$, ${\tt p\_u}$, and ${\tt p\_RO}$. ${\tt p\_T}$ models the target capacity of an agent  ({\tt pnode}) imposing a limit on the number of items (${\tt vnode}$s) that an agent can bid on.
${\tt p\_u}$ models the utility function, that can be sub-modular or not.
\footnote{In the definition of relation {\tt pcp}, the keyword {\tt one} refers to each physical node being in relation with exactly {\tt one} integer, $e.g.$, the physical capacity. Similarly, the relation {\tt pconnections} models the fact that in a network, each node is connected to {\tt some} other nodes.}

The virtual network mapping problem maps  constrained virtual networks on a constrained physical network, eventually owned by multiple, federated infrastructure providers.  Our model can be extended to capture any constraints in the form of an Alloy {\tt fact}.  As a representative example, we show in this paper how to model the {\tt fact} that physical nodes can bid on virtual nodes only if they have enough physical capacity to host them:

\vspace{-1.5 mm}
{\scriptsize
\begin{verbatim}
    fact pcapacity{
         all p: pnode | (sum vnode.(p.initBids) ) <= p.pcp
    }
\end{verbatim}
}
\vspace{-1.5 mm}

In Alloy relations are modeled by ordered tuples; this means that unordered relations must be explicit, $e.g$, our \texttt{pconnectivity} fact shows how an undirected link has to be modeled using two (directed) relations:

\vspace{-1.5 mm}
{\scriptsize
\begin{verbatim}
    fact pconnectivity{
       all disj pn1,pn2:pnode | (pn1.pid != pn2.pid) and
       (pn1 in pn2.pconnections <=> pn2 in pn1.pconnections)
     }
\end{verbatim}
}
\vspace{-1.5 mm}

Our static model includes several other facts that regulate basic networking properties. The full Allow model code can be downloaded at~\cite{BUalloy}.



\noindent
{\bf Dynamic Model.} The dynamic behavior of the network is modeled as a {\it transition system}, and the sequence of state changes is regulated by the MCA protocol.
Network states are captured in our model  using the following signature:

\vspace{-2 mm}
{\scriptsize
\begin{verbatim}
    sig netState {
        bidVectors: some bidVector,
        time: one Int,				
        buffMsgs: set message }
\end{verbatim}
}
\vspace{-1.8 mm}

The state of the physical network  is updated as bid messages are exchanged among agents (or physical nodes.)
The {\tt bidVectors} relation contains the current view of each agent, $i.e.$, the vectors ${\bf a}$, ${\bf b}$, ${\bf t}$, and ${\bf m}$ (defined in Section~\ref{sec:max-consensus} and depicted in Figure~\ref{example}.)
The relation {\tt time} models the time generation of each state, while the set of unprocessed messages is modeled with the \texttt{buffMsgs} relation. This relation captures the correspondence between states and the buffer of messages in transit. The signature message is modeled as follows:
%

\vspace{-1.5 mm}
{\scriptsize
\begin{verbatim}
    sig message{
        msgSender: one pnode,
        msgReceiver: one pnode,
        msgWinners: vnode->(pnode + NULL),
        msgBids: vnode->Int,
        msgBidTimes: vnode->Int
    }
\end{verbatim}
}
\vspace{-1.5 mm}

Aside from defining the sender and the receiver physical node, the bid \texttt{message} signature contains: the view of the sender about the maximum bid known so far on every virtual node (\texttt{msgBids}), their winners (\texttt{msgWinners}), and the time at which the highest bids were generated (\texttt{msgBidTimes}.)
%
%
Note how, when a message is being processed, these relations are used to update the states  ${\bf a}$, ${\bf b}$, ${\bf t}$, and the bundle vector ${\bf m}$ for each physical node.


%
%
%
%
%
%
%

The core of the MCA protocol is modeled by some constraint paragraphs.
In particular, the Alloy fact {\tt stateTransition} models the sequence of message processing, and the transitions from state $s$ to $s^{\prime}$:

\vspace{-1.8 mm}
{\scriptsize
    \begin{verbatim}
    fact stateTransition{
         all s: netState, s': s.next | one m:message |
         messageProcessing[s, s', m]
    }
\end{verbatim}
}
\vspace{-1.5 mm}

Using the built-in library {\tt ordering}, we can model the states of the system as an ordered sequence which keyword {\tt next} in \texttt{s.next} represents the state subsequent to \texttt{s} in the transition.

\noindent
{\bf Abstractions Efficiency.} The model we have presented so far, contains integer variables and ternary relations; see $e.g.$, the three signatures \texttt{pnode}, \texttt{bidVector} and \texttt{message}.
Ternary relations and integers were introduced for the sake of explaining our model, but lead to inefficiencies of the Alloy Analyzer.
Using such elements, our model containing the conflict resolution table of the asynchronous MCA protocol generated over $259K$ SAT clauses, for a scope as limited as $3$ physical nodes and $2$  virtual nodes. 

We obtained a more efficient model by ($i$) replacing each ternary relation with two binary relations, and by ($ii$) defining our own {\it values} ---combinations of signatures and facts--- instead of using integers ---predefined and more complex abstractions in Alloy.
As an example of signature introduced to reduce the complexity of the ternary abstractions, we show \texttt{bidTriple}:

\vspace{-1.5 mm}
{\scriptsize
\begin{verbatim}
    sig bidTriple{
        bid_v: one vnode,
        bid_b: one Int,
        bid_t: one Int,
        bid_w: one (pnode + NULL)
    }
\end{verbatim}
}
\vspace{-1.5 mm}

\noindent
To avoid using the Alloy's predefined integers (signature \texttt{Int}) we model natural numbers with the 
signature \texttt{value}:

\vspace{-1.8 mm}
{\scriptsize
\begin{verbatim}
    sig value{
        succ: set value,
        pre: set value
    }
\end{verbatim}
}
\vspace{-1.5 mm}

\noindent
Each instance of the signature \texttt{value} only models relations between numbers.
 %
%
%
Using the two relations \texttt{succ} and \texttt{pre} we model binary operators $<$, $\leq$, $>$ and $\geq$, respectively, using the binary predicates \texttt{valL[,]},
\texttt{valLE[,]}, \texttt{valG[,]} and \texttt{valGE[,]}.
For two instances \texttt{v1} and \texttt{v2} of the signature \texttt{value},
we model the inequality \texttt{v1} $\leq$ \texttt{v2} with the predicate \texttt{valLE[v1, v2]} (which in our Alloy model is equivalent to \texttt{v1 in v2.pre}).


Using these more efficient abstractions, for the same scope, we were able to reduce  the number of SAT clauses from circa $259K$ to circa $190K$, reducing the running time of our consensus assertion from circa a day to less than two hours.~\footnote{Our experiments were carried out on a Linux machine running Intel core i3 CPU at 1.4GHz and 4 GB of memory.}

\vspace{-0.05in}
\section{Using Alloy to Analyze MCA Convergence}\label{sec:property-checking}

Our model enables the study of the convergence properties of the MCA protocol. In this section we first introduce the assertion for checking such convergence property, and then we show how specific combinations of the MCA policy instantiations may or may not lead to convergence.

%
Checking the convergence property in Alloy means checking whether or not the consensus assertion holds. All the agents (physical nodes) need to reach an agreement on  ($i$) the assignment vector,
containing the identity of the winner agents (virtual nodes), ($ii$) and the bid vector.
The assertion consensus is coded in Alloy as follows~\footnote{The model presented in this paper is a simplified version of the full model described  at~\cite{BUalloy}.}:

\vspace{-1.5 mm}
{\scriptsize
\begin{verbatim}
    assert consensus{
	       (#(netState) >= val) implies consensusPred[]
    }
    pred consensusPred{
       some s: netState | all disj bv1,bv2: s.bidVectors |
        (
            (bv1.winners    = bv2.winners) and
            (bv1.winnerBids = bv2.winnerBids)
        )
    }
\end{verbatim}
}
\vspace{-1.5 mm}
%

\noindent

From the consensus literature~\cite{consensusBook}, and from previous studies on the MCA~\cite{cbba,CADpaper}, we know that the number of messages required to reach consensus is upper bounded by $D \cdot |V_H|$ where $|V_H|$ is the size of the item set, and $D$ is the diameter of the network of agents. Intuitively, this is because the maximum bid for each item, only has to traverse the network of agents once. We use this bound to set our {\tt val} parameter in the consensus assertion: after {\tt val} number of messages is being processed, a max-consensus on the bid has to be achieved.

   \begin{figure}[t]
\begin{center}
    \includegraphics[scale=0.43]{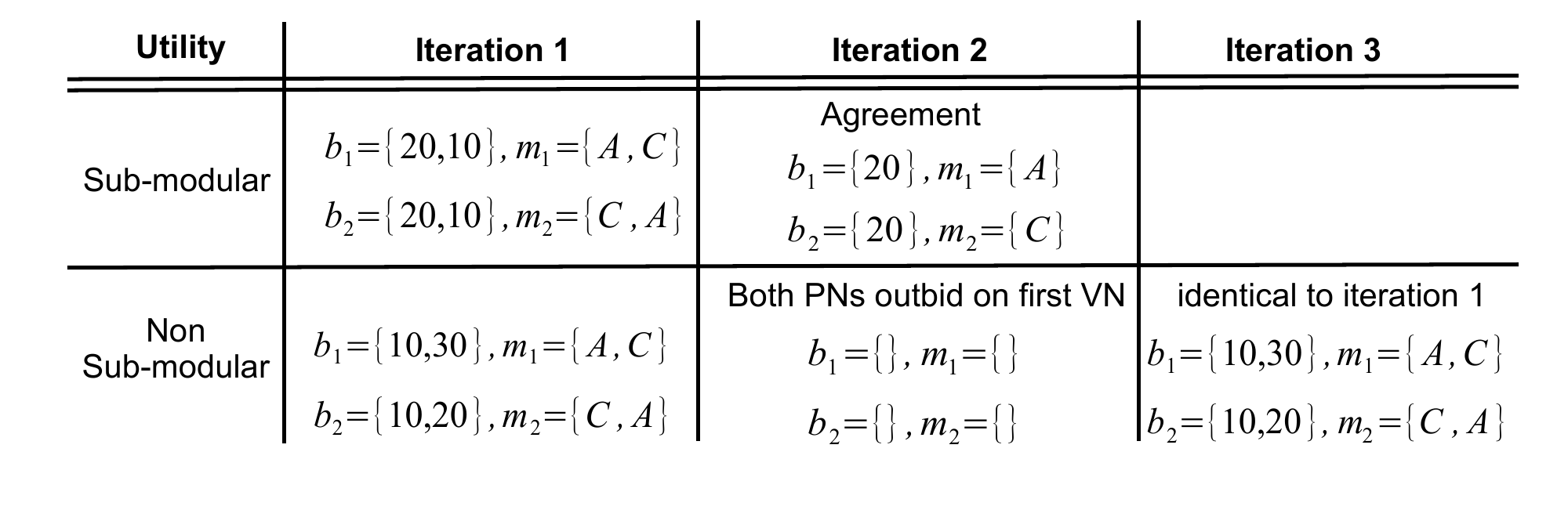}
\end{center}
\vspace{-0.2in}
\caption{The policy of releasing outbid items, combined with the non sub-modularity policy lead to instability of the MCA: with non sub-modular utility, after the first round both agents have been outbid on the first item, and their bids on the second item have been invalidated. Even bids subsequent to an outbid (see Remark 2) are released seeking a Pareto optimality. }
\label{fig:sub-modular}
\vspace{-0.25in}
\end{figure}

\noindent
{\bf  Result 1:} We checked the assertion {\tt consensus} over several scopes, for a key representative combinations of policies. {\it We found that MCA always reaches consensus, except when the utility function policy ${\tt p\_u}$ is set to non sub-modular, and the agents release (and rebid) all subsequent items to an outbid item $i.e.$, the ${\tt p\_RO}$ policy is set to true.}

To understand why the MCA protocol fails for this combination of policies, consider the scenario in Figure~\ref{fig:sub-modular} (first row):  the agent's bids do not increase as items are added to the bundle, as bids  have been generated using a sub-modular function. After exchanging the bids, item $C$ is won by agent $2$, and item $A$ is won by agent $1$. When instead MCA uses a non sub-modular function (as in Figure~\ref{fig:sub-modular} second raw), bids can increase as items are added to the bundle, and releasing items (subsequent to an outbid node to refresh their bids) causes oscillations, and hence the MCA failure to reach a conflict-free assignment.



\noindent
{\bf Result 2:} We also tested the consensus property under circumstances of protocol misbehavior or misconfiguration. In particular,  we removed from our model the necessary condition discussed in {\it Remark~\ref{rm:overbid}}, allowing physical nodes to re-bid after they were outbid on a virtual node and, as expected, we found instances in which, consensus (a conflict-free assignment) is not reached. 
A consequence of this sanity-check  for our model is that the MCA protocol is not resilient to {\it rebidding attacks}, $i.e.$, malicious agents can perform a denial of service attack by rebidding even on outbid items.
\footnote{
A thorough analysis on how to design and implement solutions to detect or prevent malicious MCA agents is left as an open research question. However, singular malicious user behavior can be isolated by requiring every agent to sign their messages before broadcasting, using a unique ID. By keeping track of the bidding history of their first hop neighborhood, agents could then detect rebidding attacks (condition in Remark~\ref{rm:overbid}), ignoring subsequent invalid bid messages.}
%




\vspace{-4mm}
\section{Related Work}
    \label{sec:related-work}
    \noindent
{\bf Protocol Verification with Alloy.} Theorem proving and model checking tools have been widely used to analyze and verify (distributed) algorithms and protocols~\cite{bhargavan2002formal, havelund1996experiments,zave2009lightweight,sadeghian2013formal} for a wide range of (networked) applications, as they allow with minimal implementation efforts to verify complicated properties.
The attention towards {\it lightweight} model-finding tools~\cite{Holzmann:2003:SMC:1405716, Jackson:2006}, as an alternative
to model checking tools~\cite{yovine1997kronos, groote2007formal} has only recently increased, due to the ease of use, and to the automation that they have introduced. We only cite a representative set of references to define our work in context.
In~\cite{zave2009lightweight} and \cite{sadeghian2013formal}, the authors 
study with an Alloy model Chord~\cite{ionstoica1chord}, a peer-to-peer distributed hash table  protocol.
%
\junk{
In \cite{sadeghian2013formal}, the authors use Alloy to verify some of the
Chord protocol's properties. In a peer-to-peer network, nodes can have the same ID.
In this work the network and the Chord protocol are modeled using Alloy and
different features of Pure-join model of Chord protocol are investigated under the
condition that different nodes can have the same ID.
Alloy facts help in order apply such condition and using assertions, properties of the model can be checked and verified.
}

Alloy has been applied to model and study the properties of other protocols as well~\cite{arye2011toward,chen2006design, smirzaeiOpenFlow}. In \cite{arye2011toward} for example, Alloy is used to analyze the properties of the Stable Path Problem (SPP), and to verify sufficient conditions on SPP instances.

\junk{
In \cite{chen2006design} a security model for secure communication is presented.
This security model targets the messages' integrity,
confidentiality, authentication and non-repudiation.
In order to validate the correctness of the suggested security model,
Alloy language and Alloy analyzer is used.
Subsequently the logical formulas, definitions and security properties
are codified and verified using Alloy.
}

\noindent
{\bf Verifying Correctness of Networking Mechanisms.} Recent work has been also carried out to verify the correct behavior of many networking mechanisms. For example, there has been  interest in verifying that network forwarding rules match the intent specified by the administrator~\cite{VargheseNSDI2013,VeriFlow}, or even in building tools to debug the network forwarding plane in the context of Software-Defined Networks~\cite{ndb}. 

Approaches that verify the correct behavior of the routing~\cite{Mai:2011:anteater} or the forwarding~\cite{Al-Shaer:2010:FCA:1866898.1866905,Canini:2012:NWT:2228298.2228312} mechanisms have also been investigated. The authors in~\cite{Mai:2011:anteater} for example, propose via SAT instances to statically analyze the router configurations of the data plane, to check isolation errors and network disconnections caused by misconfigurations.
\junk{
FlowChecker~\cite{Al-Shaer:2010:FCA:1866898.1866905} tries to find intra-switch misconfiguration in OpenFlow switches~\cite{McKeown:2008:OEI:1355734.1355746}.
This tool translates FlowTable configurations into
boolean expressions using Binary Decision diagrams (BDD)
and provides a property-based verification interface using
BDD-based symbolic model checking and temporal logic.
Using model checking techniques, beside FlowChecker, many other tools on the analysis of OpenFlow have been introduced.
NICE~\cite{Canini:2012:NWT:2228298.2228312} is a controller verification
tool which uses model checking and symbolic execution to
automate testing Openflow applications. NICE attempts to explore
the state space of the whole network and find the invalid
system states.
In contrast to our work which is a solution for verifying a protocol's properties regardless of the individual instances of networks, FlowChecker and other tools (such as Anteater, NICE and \emph{etc.}) provide the network analysts with the ability to check the network properties after deployment.
In model checking techniques, usually systems are modeled as transition system
(as in our model using signature \texttt{netState} and message processing predicates).
In~\cite{5339690-al-Shaer} the whole network is modeled as a finite state machine.
The goal in this work is to check correctness of network reachability per packet.
The temporal properties of the network and state of the packets are specified using
Computation Tree Logic (CTL) and analyzed facilitating
BBDs.
}

Similar to all these approaches, our work also aims to verify the correctness of a network mechanism, but our focus is on the Max Consensus Auction; in particular, on the virtual network mapping, a management application that infrastructure providers use {\it during} the creation of a virtual network, not after a (virtual) network has been instantiated.

\vspace{-1mm}
\section{Conclusions}
    \label{sec:conclusions}

\junk{
1. Max-consensus auction (MCA) protocols are amazing, useful and interesting, look at all these applications, here is how MCA works...
2. Be careful if you use MCA because a wrong policy combination can cause instability (oscillations) i.e. no convergence to a conflict free assignment.(we have to define convergence)
3. We prove with an Alloy that a combination of policies causes instability (submmodular + release)
4. We give you an Alloy model to test your own MCA policies
}

Max Consensus-based Auction protocols are a recent solution that allows a set of communicating agents to obtain a conflict-free (distributed) allocation of a set of items, given a common network utility maximization goal. We extracted the common mechanisms of such protocols, renaming them MCA: a bidding mechanism, where agents independently bid on a single or on multiple items,  and an agreement (consensus) mechanism, where agents exchange their bids for a distributed winner determination.  Each MCA mechanism can be instantiated with a wide-range of policies that lead to different behaviors and protocol properties.

In this paper, we used the Alloy Language to model the MCA protocol, and verify its convergence properties under a range of different policies. Our MCA model is application agnostic, but we described our result in context of the virtual network mapping problem. With our model, we were able to show how given combination of MCA policies lead to instability (oscillations) $i.e.$, no convergence to a conflict free assignment is guaranteed, and that MCA is not immune to denial of service attacks as {\it rebidding attack}.
Our released Alloy model can be used to verify the correctness of the MCA protocol, for a wide range of policies and applications,  or extended to include property-checking features for large instance of the model.


\setstretch{0.9}
\bibliographystyle{abbrv}
\vspace{-1mm}
\bibliography{allBibs}

\begin{thebibliography}{10}

\bibitem{BUalloy}
{MCA Alloy model code.} \url{http://csr.bu.edu/alloy/}.

\bibitem{Al-Shaer:2010:FCA:1866898.1866905}
{Al-Shaer, E. and Al-Haj, S.}
\newblock Flowchecker: Configuration analysis and verification of federated
  openflow infrastructures.
\newblock In {\em Proc. of SafeConfig}, pages 37--44, New York, NY, USA, 2010.
  ACM.

\bibitem{arye2011toward}
M.~Arye, R.~Harrison, R.~Wang, P.~Zave, and J.~Rexford.
\newblock Toward a lightweight model of {{BGP}} safety.
\newblock {\em Proc. of WRiPE}, 2011.

\bibitem{bhargavan2002formal}
K.~Bhargavan, D.~Obradovic, and C.~A. Gunter.
\newblock Formal verification of standards for distance vector routing
  protocols.
\newblock {\em Journal of the ACM (JACM)}, 49(4):538--576, 2002.

\bibitem{MCA-grid}
{Binetti, G. $et$ $al$}.
\newblock A distributed auction-based algorithm for the nonconvex economic
  dispatch problem.
\newblock {\em Industrial Informatics, IEEE Trans. on}, 10(2):1124--1132, May
  2014.

\bibitem{Canini:2012:NWT:2228298.2228312}
{Canini, M. $et$ $al$}.
\newblock {A {NICE} Way to Test Openflow Applications}.
\newblock In {\em Proc. of NSDI}, pages 10--10, Berkeley, CA, USA, 2012.

\bibitem{chen2006design}
C.~Chen, P.~Grisham, S.~Khurshid, and D.~Perry.
\newblock Design and validation of a general security model with the alloy
  analyzer.
\newblock In {\em Proc. of the ACM SIGSOFT}, pages 38--47, 2006.

\bibitem{cbba}
{Choi, Han-Lim $et$ $al$}.
\newblock Consensus-based decentralized auctions for robust task allocation.
\newblock {\em IEEE Trans. on Robotics}, Aug 2009.

\bibitem{cook1971complexity}
S.~A. Cook.
\newblock The complexity of theorem-proving procedures.
\newblock In {\em Proc. of ACM Theory of comp.}, pages 151--158, 1971.

\bibitem{CADpaper}
F.~Esposito, D.~{Di Paola}, and I.~Matta.
\newblock On distributed virtual network embedding with guarantees.
\newblock {\em ACM/IEEE Transactions on Networking. Accepted (to appear)}, Nov
  2014.

\bibitem{groote2007formal}
{Groote, J. F. $et$ $al$}.
\newblock {\em The formal specification language mCRL2}.

\bibitem{ndb}
{Handigol, Nikhil $et$ $al$}.
\newblock Where is the debugger for my software-defined network?
\newblock HotSDN '12, pages 55--60, 2012.

\bibitem{havelund1996experiments}
K.~Havelund and N.~Shankar.
\newblock Experiments in theorem proving and model checking for protocol
  verification.
\newblock In {\em FME'96}, pages 662--681. Springer, 1996.

\bibitem{Holzmann:2003:SMC:1405716}
G.~Holzmann.
\newblock {\em Spin Model Checker, the: Primer and Reference Manual}.
\newblock Addison-Wesley Professional, first edition, 2003.

\bibitem{ionstoica1chord}
{I. Stoica $et$ $al$}.
\newblock Chord: A scalable peer-to-peer lookup service for internet
  applications.
\newblock In {\em SIGCOMM'01}, pages 27--31.

\bibitem{Jackson:2006}
D.~Jackson.
\newblock {\em Software Abstractions: Logic, Language, and Analysis}.
\newblock The MIT Press, 2006.

\bibitem{VargheseNSDI2013}
{Kazemian, Kazemian $et$ $al$}.
\newblock Real time network policy checking using header space analysis.
\newblock In {\em Proc. of the 10th USENIX}, nsdi'13, pages 99--112, Berkeley,
  CA, USA, 2013.

\bibitem{VeriFlow}
{Khurshid, Ahmed $et$ $al$}.
\newblock Veriflow: Verifying network-wide invariants in real time.
\newblock In {\em Proc. of HotSDN '12}, pages 49--54, NY, USA, 2012.

\bibitem{consensusBook}
N.~A. Lynch.
\newblock {\em Distributed algorithms}.
\newblock Morgan Kaufmann, 1996.

\bibitem{Mai:2011:anteater}
{Mai, Haohui $et$ $al$}.
\newblock Debugging the data plane with anteater.
\newblock In {\em Proc. of the ACM SIGCOMM '11}, pages 290--301, NY, USA, 2011.
  ACM.

\bibitem{smirzaeiOpenFlow}
S.~Mirzaei, S.~Bahargam, R.~Skowyra, Kfoury, a., and A.~Bestavros.
\newblock Using alloy to formally model and reason about an openflow network
  switch.
\newblock {\em CS Dept., Boston University, Tech. Rep. 2013-007}, 2013.

\bibitem{karp}
{R.M. Karp}.
\newblock {Complexity of Computer Computations}.
\newblock In {\em Reducibility Among Combinatorial Problems, Miller and
  Thatcher.}, 1972.

\bibitem{sadeghian2013formal}
H.~Sadeghian, A.~Samadi, and H.~Haghighi.
\newblock Formal analysis of pure-join model of chord using alloy.
\newblock In {\em ICSESS}, May 2013.

\bibitem{torlak2007kodkod}
E.~Torlak and D.~Jackson.
\newblock Kodkod: A relational model finder.
\newblock In {\em Tools and Algorithms for the Construction and Analysis of
  Systems}, pages 632--647. Springer, 2007.

\bibitem{yovine1997kronos}
S.~Yovine.
\newblock Kronos: A verification tool for real-time systems.
\newblock {\em International Journal on Software Tools for Technology
  Transfer}, pages 123--133, 1997.

\bibitem{zave2009lightweight}
P.~Zave.
\newblock Lightweight verification of network protocols: The case of chord.
\newblock 158, 2009.

\end{thebibliography}

\end{document}